\documentclass[%
 reprint,
superscriptaddress,
 amsmath,amssymb,
 aps,
]{revtex4-1}

\usepackage{graphicx}
\usepackage{dcolumn}
\usepackage{bm}

\begin{document}

\preprint{APS/123-QED}

\title{Polarisation drift compensation in an 8 km long Mach-Zehnder fibre-optical interferometer for quantum communication}

\author{G. B. Xavier}
\email{guix@opto.cetuc.puc-rio.br}
\affiliation{%
 Center for Telecommunication Studies, Pontifical Catholic University of Rio de Janeiro, R. Marqu\^es de S\~ao Vicente, 225 G\'avea - Rio de Janeiro - Brazil - 22451-900} 

\author{T. R. da Silva}%

\author{G. P. Tempor\~ao}

\author{J. P. von der Weid}



\begin{abstract}
We experimentally stabilise the polarisation drift between the arms of an 8 km-long fibre-optical Mach-Zehnder interferometer, while simultaneously compensating the phase fluctuations. The single photons are wavelength-multiplexed with three classical channels, which are used as feedback for the control systems. Two of these channels are used for the active polarisation control systems, while the other is used to phase-lock the interferometer. We demonstrate long-term stabilisation of the single-photon visibility when polarisation control is used.
\end{abstract}

\maketitle


\section{\label{sec:level1}Introduction}

Quantum communication \cite{Gisin_Nat_Photon} has evolved considerably in the last years, as research in this new field has widened extensively. In particular, an interest in long-distance interferometry with single photons has recently increased, with applications in quantum cryptography \cite{Noh_PRL} and quantum repeaters \cite{Geneva_repeater}. Recently, a 6 km long fibre-optical phase-stabilised Michelson interferometer for single-photons has been demonstrated \cite{Noh_Opex}. It employed Faraday mirrors to compensate for the polarisation drift in the optical fibres. In a Mach-Zehnder (MZ) configuration, independent polarisation drift in the two arms will lead to a fluctuation of the single-photon visibility over time, and an active solution is needed, like in point-to-point optical links for polarisation encoded quantum communication \cite{Gisin_RMP}.
In this letter we demonstrate that it is possible to maintain the single-photon visibility stable over a long period of time, when using active polarisation control in both arms of the interferometer. We employ two wavelength multiplexed feedback signals using continuous-wave (CW) lasers \cite{Guix_NJP}. An additional CW laser, at a different wavelength, is used to lock the phase difference between the two arms of the interferometer, using a piezo-electric fibre stretcher to compensate the phase drift. The frequency of this drift can even reach a few kHz depending on the interferometer length, due to environmental factors such as temperature and mechanical vibrations \cite{Minar_PRA}. We are nevertheless able to compensate this phase drift while simultaneously controlling the polarisation drift in the interferometer arms.

\section{Experimental Setup}

The experimental setup is shown in Fig. 1, with four lasers being combined in a dense wavelength division multiplexer (DWDM) using the standard 100 GHz adjacent channel spacing (0.8 nm @ 1550 nm), and ~ 1.6 dB insertion loss. The two lasers for the polarisation control feedback are standard telecom distributed feedback (DFB) laser diodes at $\lambda_{P1}$ = 1545.32 nm and $\lambda_{P2}$ = 1546.92 nm. Manual polarisation controllers are employed to adjust their input polarisation states to be non-orthogonal with maximum overlap \cite{Guix_NJP}. The other lasers are two tuneable external cavity lasers centred at $\lambda_Q$ = 1546.12 nm and $\lambda_{PH}$ = 1547.72 nm, with the first one (the quantum channel), used as the pseudo-single photon source. It is combined with an optical attenuator set to produce an average of 0.5 photons per detection window at the output of the DWDM. The other laser is employed as feedback for the phase stabilisation system. Before each feedback laser a band-pass filter (BPF), consisting of an optical circulator with a fibre Bragg grating (FBG) centred at each laserÕs wavelength, is used to remove the broadband unwanted spontaneous emission from the lasers, which would otherwise fall in-band with the single photons. The input power of the feedback lasers is adjusted such that there is a negligible number of Raman spontaneous scattered photons impinging in the quantum channel \cite{Guix_EL_Raman}.

\begin{figure}
\includegraphics[width=0.48\textwidth]{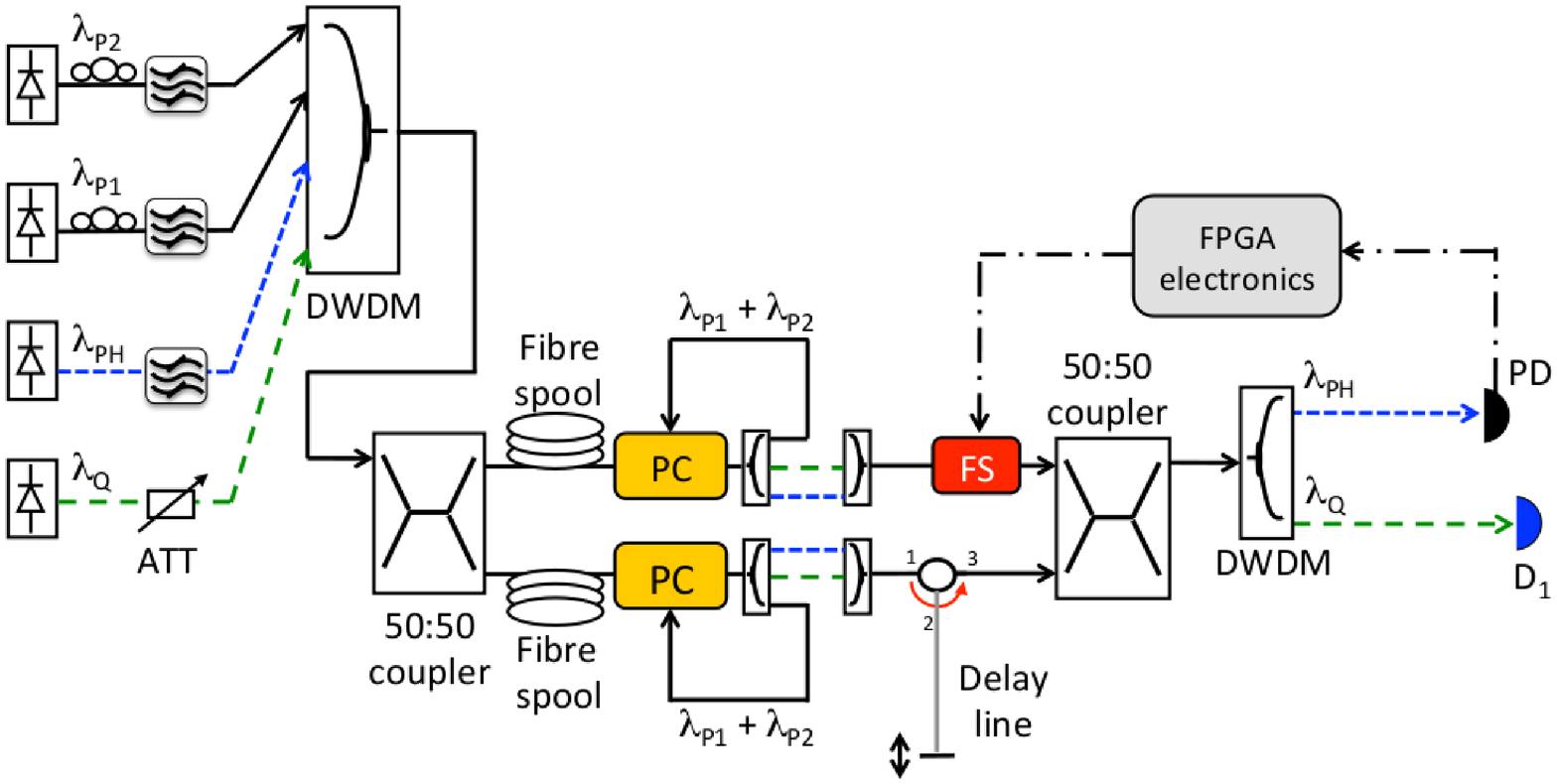}
\centering
\caption{\label{figure1}{(Color online) Experimental setup. The dashed lines are used to identify the phase control laser and the single-photons when not multiplexed with any other wavelengths. The solid grey line represent free-space in the delay line, while dashed and dotted black represent electrical connections.}}
\end{figure}

The long-distance MZ interferometer is built using two optical fibre spools consisting each of 8 km of dispersion-shifted fibre as its arms, and two 50:50 fibre couplers as shown in Fig. 1. In each arm we have an active polarisation control system (PC) placed after the fibre spools, whose function is to keep the polarisation state in that arm stable. After each PC we place another DWDM, identical to the one used to combine the four lasers, to remove the two feedback polarisation signals, detect them and then allow the polarisation control algorithm to stabilise the polarisation state. The necessary components to perform this (linear polarisers, manual polarisation controllers, p-i-n detectors, electronics, etc) are not shown for the sake of clarity, and details on the control system can be found in \cite{Guix_NJP}.

The single photons and the phase reference signal are split for a few centimetres and then recombined in another DWDM, such that the single photons are not phase-locked for a short distance. This introduces a long-term phase drift allowing us to obtain the single-photon interference curves as a function of time, without having to use an extra actuator to change the phase difference between the arms of the interferometer. Following the recombination of the $\lambda_Q$ and $\lambda_{PH}$ signals, a piezo-electric based fibre stretcher (FS) is placed in one of the arms and used to rapidly change the length of the fibre up to a couple of mm, therefore changing the phase difference between the arms. At one of the output ports of the second 50:50 fibre coupler, a DWDM splits $\lambda_{Q}$ from $\lambda_{PH}$, with the later detected by a p-i-n photodiode (PD), then measured with a field programmable gate array (FPGA), which processes the information and performs the control algorithm. The electrical output of the FPGA is then amplified and fed back to the fibre stretcher closing the control loop. The single-photons exit the DWDM through the other output and an extra BPF (with a FBG centred at $\lambda_{Q}$) is used to remove unwanted photons that leaked from cross-talk in the DWDM. A single photon counting module (SPCM) working in gated mode, $D_1$, with 15\% quantum efficiency, is used to detect the single photons, with a gate repetition frequency of 100 kHz and a measured dark count probability of 3.2 x $10^{-5}$ per gate of 2.5 ns. A free-space optical delay line composed of a movable mirror with a fibre connector mounted on a collimator, combined with a circulator with a total loss of  ~ 4 dB, is used to coarsely adjust the total path length difference between the two arms. A low coherence light source (a LED) is used in place of the phase control laser for this adjustment. We are able to adjust the path length difference within $\sim$ 1 mm with this technique.

\section{Results and discussions}

We record the net single-photon counts (subtracting only the dark counts) over $D_1$ using a 1 s integration time, with the polarisation control turned on and then off for $\sim$ 85 minutes in each run. These results are displayed in Figs. 2a and 2b. The phase control system is also turned on during these measurements, and it is why we are able to observe interference fringes. In Fig. 2c classical measurements of the optical intensity of $\lambda_{PH}$ are shown as measured by detector PD, with the phase control turned on and off. We clearly see that the optical intensity indeed fluctuates in the ms time scale [7] when phase control is off, therefore the SPCM would average out the single-photon interference curves with a 1 s integration time. We observe from Figs. 2a and 2b that, with polarisation control active, the optical visibility of the fringes remain constant, while when it is turned off, the drift causes the visibility to change, due to a less than optimal overlap between the polarisation states at the end of the two arms. 

\begin{figure}
\includegraphics[width=0.48\textwidth]{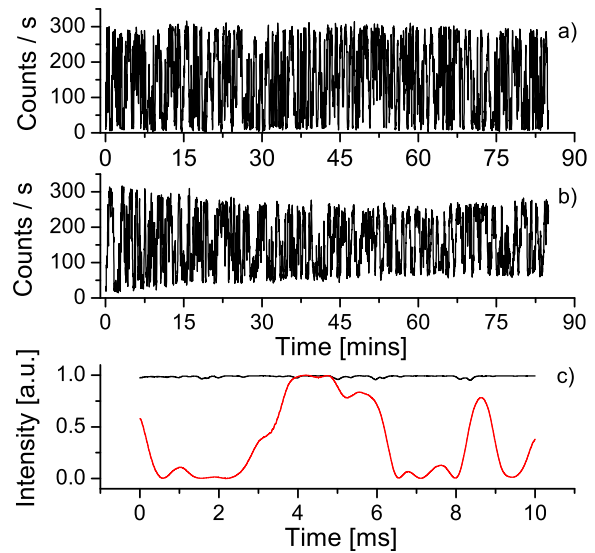}
\centering
\caption{\label{figure1}{(Color online) Experimental setup. The dashed lines are used to identify the phase control laser and the single-photons. a) Single-photon interference at $D_1$ with both polarisation and phase control turned on. The slow fluctuation is due to the separation of $\lambda_Q$ and $\lambda_{PH}$  after each PC (see text for details). b) identical as a) but with polarisation control deactivated. c) classical interference at PD with phase control on (black) and off (red).when not multiplexed with any other wavelengths. The solid grey line represent free-space in the delay line, while dashed and dotted black represent electrical connections.}}
\end{figure}

We calculate the envelope of the interference fringes in Figs. 2a and 2b, and from it we can work out the maximum possible visibility at each measured point. The single-photon visibility is defined as   $V=|(C_{\text{max}} - C_{\text{min}})/(C_{\text{max}}+C_{\text{min}})|$ where $C_{\text{max}}$ are the counts in the upper bound of the envelope and $C_{\text{min}}$ the counts in the lower bound. The results for the visibility are arranged in a histogram and plotted in Fig. 3. We clearly observe the much larger spread of the distribution for the case with the polarisation control off, and this is clearly reflected from Fig. 2b. The average net visibilities are 92.6 $\pm$ 0.14 \% and 71.0 $\pm$ 8.1 \% for the cases of polarisation control on and off respectively, showing that without control the polarisation drift lowers the average visibility as well as allowing it to wander considerably. At 92.6 \% the visibility is still sufficient to allow quantum communication protocols to take place [5], whereas 71.0 \% is too low for most known applications. 

\begin{figure}
\includegraphics[width=0.48\textwidth]{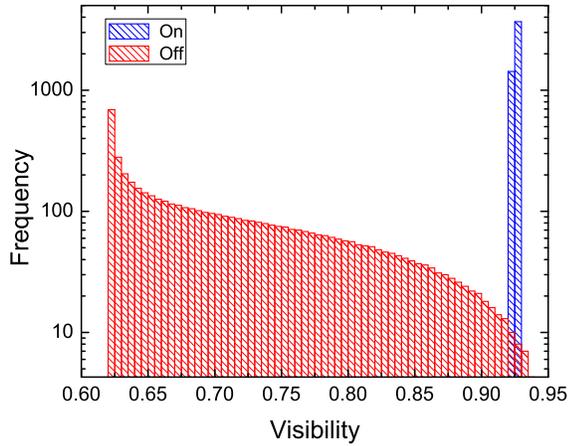}
\centering
\caption{\label{figure1}{(Color online) Histogram of the calculated visibilities from the envelopes of the single-photon interference curves in Fig. 2a and b.}}
\end{figure}

\section{Conclusions}

We have experimentally demonstrated stable visibility of single-photon interference in an 8 km long fibre-optical MZ interferometer. This stability was provided by actively controlling the polarisation state of the single-photons in each arm of the interferometer. The phase was also actively stabilised, allowing us to observe single-photon interference fringes with 92.6 \% net visibility

 \begin{acknowledgments}
The authors wish to acknowledge financial support from CAPES, Faperj and CNPq.
\end{acknowledgments}



\end{document}